\documentclass{article}
\usepackage[utf8]{inputenc}
\usepackage{arxiv}
\usepackage{amsmath}
\usepackage{enumerate}
\usepackage{subfigure}
\usepackage{graphicx}
\usepackage{booktabs}
\usepackage{url}
\usepackage{natbib}

\graphicspath{{.}}

\usepackage[inline,shortlabels]{enumitem}

\usepackage[nameinlink]{cleveref}
\crefname{figure}{Fig.}{Figs.}
\Crefname{figure}{Figure}{Figures}
\crefname{equation}{Eq.}{Eqs.}
\Crefname{equation}{Equation}{Equations}
\crefformat{section}{#2§#1#3}
\crefname{section}{§}{§§}
\Crefname{section}{Section}{Sections}
\crefname{table}{Table}{Tables}
\crefname{appendix}{Appendix}{Appendices}

\newcommand{\eg}{e.g.,\ }

\newcommand{\ie}{i.e.,\ }

\newcommand{\centigrades}{{\textdegree{}C}}

\DeclareMathOperator*{\argmin}{\arg \min}

\title{Real-World Implementation of Reinforcement Learning Based Energy Coordination for a Cluster of Households}

\author{%
Gargya Gokhale~\thanks{Presented at the RLEM Workshop at the 10th ACM International Conference on Systems for Energy-Efficient Buildings, Cities, and Transportation (BuildSys '23), November 15--16, 2023, Istanbul, Turkey. Publication rights licensed to ACM. \\ DOI: 10.1145/3600100.3625681.}\\
IDLab, Ghent University--imec\\
Gent, Belgium \\
\texttt{gargya.gokhale@ugent.be} \\
\And
Niels Tiben\\
IDLab, Ghent University--imec\\
Gent, Belgium \\
\And
Marie-Sophie Verwee\\
IDLab, Ghent University--imec\\
Gent, Belgium\\
\And
Manu Lahariya\\
IDLab, Ghent University--imec\\
Gent, Belgium\\
\And
Bert Claessens\\
IDLab, Ghent University--imec\\
beebop.ai \\
\And
Chris Develder\\
IDLab, Ghent University--imec\\
Gent, Belgium\\
}

\begin{document}

\maketitle

\begin{abstract}
Given its substantial contribution of 40\% to global power consumption, the built environment has received increasing attention to serve as a source of flexibility to assist the modern power grid. In that respect, previous research mainly focused on energy management of individual buildings. In contrast, in this paper, we focus on aggregated control of a set of residential buildings, to provide grid supporting services, that eventually should include ancillary services. In particular, we present a real-life pilot study that studies the effectiveness of reinforcement-learning (RL) in coordinating the power consumption of 8 residential buildings to jointly track a target power signal. Our RL approach relies solely on observed data from individual households and does not require any explicit building models or simulators, making it practical to implement and easy to scale. We show the feasibility of our proposed RL-based coordination strategy in a real-world setting. In a 4-week case study, we demonstrate a hierarchical control system, relying on an RL-based ranking system to select which households to activate flex assets from, and a real-time PI control-based power dispatch mechanism to control the selected assets. Our results demonstrate satisfactory power tracking, and the effectiveness of the RL-based ranks which are learnt in a purely data-driven manner. 
\end{abstract}


\section{Introduction}
\label{sec:intro}

The past decades have witnessed a steady increase in the integration of renewable energy sources~(RES) into the power grid, a trend that is expected to continue in the coming years. This growth in RES deployment implies a rising need for energy flexibility to ensure reliable and secure grid operation. Conventionally, energy flexibility in power grids was obtained using fast ramping energy sources such as gas-powered generators. However, with growing emphasis on clean, CO2-free energy, demand-side flexibility has been gaining significant attention, \eg grid level batteries, industrial processes. In this respect, the built environment holds a significant potential as an as of yet (largely) untapped source of flexibility.
Indeed, buildings account for over 40\% of global energy consumption~\cite{built_env_con}, and offer ample opportunities to manage their consumption flexibly.
The latter typically entails using their intrinsic thermal mass as a passive storage element and shifting their consumption patterns whilst satisfying necessary comfort levels~\cite{mpc_rl_review}.

While buildings constitute a huge potential for contributing towards demand side flexibility, realizing a \emph{control framework} capable of exploiting this flexibility is a highly challenging problem~\cite{energy_flex}.
This is particularly due to the non-linear dynamics of the building and the stochastic operating conditions related to their dependence on outside temperatures, human behavior, solar irradiation, etc. 
Clearly, the challenge is exacerbated when extended to coordination of a group of buildings or an entire neighborhood.
Consequently, most works in the well-researched area of thermal and HVAC control for buildings (see \cite{mpc_review_2023, rl_review} for recent overviews) focus on cost or power consumption minimization for a \emph{single building}.

Those works adopt advanced control techniques to develop building-level controllers, with \emph{Model Predictive Control (MPC)}~\cite{mpc-basics} being a predominantly used strategy.
Works such as~\cite{swiss_mpc_frequency,wang_mpc_field_2023, mpc_brussels} have demonstrated the deployment of MPC-based control in large commercial buildings and have shown significant improvement in energy flexibility of these buildings as compared to business-as-usual controllers. 
Such MPC solutions require a good quality model of the building(s) at hand, which is challenging in practice.
This is one of the reasons for limiting the control to a single, typically commercial\footnote{Since for that scale, constructing/obtaining a sufficiently accurate model is affordable.} building rather than, \eg a cluster.
Given that such lack of scalability in MPC-based control forms a major bottleneck, recent research is moving towards data-driven control paradigms such as Reinforcement Learning~(RL).

Being data-driven, \emph{RL-based control} relies on interacting with the building and iteratively improving the control policy.
Spurred by advances in deep learning and compute capabilities, prior works have implemented different RL-based control strategies for optimization in individual buildings~\cite{rl_hvac_2020, rl_real_1}, a group of buildings~\cite{tl_rl_buildings, merlin} and for coordination of a cluster buildings~\cite{coordinated_DRL, multi-agent_2023, marlisa}. While these works show the performance benefits of using RL-based controllers for different buildings and optimization objectives, most works are limited to simulation environments such as BOPTEST, CityLearn~\cite{boptest, citylearn} or utilize a high-fidelity building model based on EnergyPlus or Modelica~\cite{energyplus, modelica}. Additionally, prior works on grid interactive buildings such as~\cite{coordinated_DRL, multi-agent_2023, marlisa}, primarily consider explicit storage assets such as electric vehicles~(EVs), batteries or hot water storage tanks for flexibility. While effective, such storage assets need external installation and carry an investment cost. However, utilizing passive thermal storage in buildings (building thermal mass)~\cite{thermal_mass} presents an (almost) cost-free, thermal storage alternative, a flexibility element that has often been overlooked in past works related to building coordination. 

Our pilot study aims to contribute to this lesser-researched area of energy coordination in residential buildings, utilizing only their intrinsic thermal mass~ (\ie without any explicit storage element). To the best of our knowledge, this work is among the first works to present a real-world RL-based deployment of an energy coordination framework for a group of households without using any prior information or simulator models for these buildings.
The main contributions of our case study on controlling a cluster of residential housing units (\cref{sec:problem_statement}) are that we:
\begin{itemize}
    \item Implement a real-world proof-of-concept demonstration of a scalable, data-driven RL-based approach~(\cref{sec:methodology}) for coordinating a group of residential buildings;
    \item Demonstrate the power tracking capability of the group of residential buildings for providing short-term DR services~(\cref{sec:results});
    \item Discuss the challenges encountered during this pilot study and present future research directions for unlocking DR-potential from residential buildings~(\cref{sec:conclusions}).
\end{itemize}

\section{Problem Statement}
\label{sec:problem_statement}
\subsection{Case Study}
The pilot study undertaken in this work relates to a group of residential apartments whose energy consumption is managed by a demand aggregator. Following works such as~\cite{swiss_mpc_frequency}, we assume that the aggregator offers a power tracking service~(similar to secondary frequency control) over a short time interval~(sub-hour), when required.
During normal operation (when no DR service is needed), the aggregator is responsible for maintaining the indoor room temperature in each household within the user defined comfort bounds.
These comfort bounds refer to a narrow band of $\pm$ 1{\centigrades} around a target room temperature 
that is set by each user.
This normal operation represents the business-as-usual~(BAU) scenario and the aggregator aims to maximize the comfort of the users by maintaining the room temperature as close to the target setpoint as possible.  

However, in case of a DR event, the aggregator must switch from this BAU scenario and follow the required power signal requested by the system operator.
To do so, the aggregator must adjust the power consumption of different households, to ensure that the aggregate power consumption of all households  together follows the requested tracking signal.
This entails that some of the houses need to deviate from their BAU scenario --- either being heated slightly above the setpoint (power upregulation) or being allowed to cool down slightly below the requested setpoint (power downregulation) --- whilst ensuring user comfort~($\pm$ 1{\centigrades} around the target temperature setpoint).
Thus, during a DR event, the aggregator 
needs to identify suitable (flexible) households that must deviate from the BAU scenario and by how much.

\subsection{Problem Formulation}
This section formulates the energy coordination problem from the perspective of the demand aggregator.
First, we model each household as a partially observable Markov Decision Process~(MDP). MDP is a commonly used framework for modelling sequential decision-making problems and is characterized by a state space~$\mathbf{X}$,  an action space~$\mathbf{U}$, a state transition function~$f$ and a cost function~$\rho$~\cite{sutton-barto}.
In our case, the \emph{state} for house $h$ is defined as $\textbf{x}^{h}_t = \{t, T^{h}_{o,t}, T^{h}_{r,t-k}, \ldots, T^{h}_{r,t}\}$, \ie the time-of-day~$t$, the past $k$ room temperature values, the current room temperature~$T^{h}_{r,t}$ and outside air temperature~$T^{h}_{o,t}$.
An \emph{action}~$u^{h}_t$ is the heating power (ON or OFF), and the \emph{cost} function is the amount of energy consumed~($G^{h}_t(\textbf{x}^{h}_{t}, u^{h}_t)$). The transition function~($f(\textbf{x}^{h}_{t}, u^{h}_t)$) relates to the natural evolution of room temperature over time~(duration of each time step is 15 minutes).
Note that the actual system cannot be observed fully, with essential quantities such as building thermal mass being hidden, making it a partially observable MDP.
To mitigate this lack of complete information, we use a (short) history of past room temperature values over the last $k$ time steps~\cite{sutton-barto}.
For the BAU scenario, the control policy~($\pi^{\text{h}}_{\text{b}}$) used for maintaining the indoor temperature within user comfort levels is given by:
\begin{equation}
    \pi^{h}_{\text{b}}(\textbf{x}^{h}_{t}) = \begin{cases}
        0       &:      T^{h}_{r,t} > T^{h, \text{set}}_{r,t}\\
        1       &:      T^{h}_{r,t} \leq T^{h, \text{set}}_{r,t}\\
    \end{cases}
\label{eq:bau_policy}
\end{equation}
where $T^{h, \text{set}}_{r,t}$ denotes the user defined target temperature for time step $t$. 
We define the aggregate energy consumption~($G$) of the entire portfolio of households during time step $t$ as:
\begin{equation}
    G(\textbf{x}_t,\textbf{u}_t) = \sum_{h=1}^{N}G^{h}_t(\textbf{x}^{h}_{t}, u^{h}_t)
\end{equation}
where, $\textbf{x}_t,\textbf{u}_t$ represent the collection of current states and actions of all the houses. 
During BAU scenario, individual actions $u^{h}_{t}$ are obtained from the respective BAU policies as defined in~\cref{eq:bau_policy}.

However, during a DR event, we assume that the system operator conveys a tracking signal~($\kappa_t$) at the beginning of the DR event~($t_{\text{d}}$), for a fixed duration $\tau$ and the aggregator seeks to match it by deviating from  BAU scenario. This is formulated as a bi-level optimization problem given by:
\begin{equation}
\begin{split}
    \min_{\textbf{u}_{t_{\text{d}}}, \ldots, \textbf{u}_{t_{\text{d}}+\tau}} &\sum_{t=t_{\text{d}}}^{t=t_{\text{d}} + \tau}|\kappa_t - G(\textbf{x}_t,\textbf{u}_t)| \\
    \text{s.t} \ \textbf{u}^{h} \in \argmin_{u^{h}_{1}, \ldots, u^{h}_{T}} &\left[\sum_{t=1}^{T}G^{h}_t(\textbf{x}^{h}_{t}, u^{h}_t) - G^{h}_t(\textbf{x}^{h}_{t}, \pi^{h}_{\text{b}}(\textbf{x}^{h}_{t}))\right] \quad \forall \ h \\
    T^{h, \text{set}}_{r,t} - 1 &\leq T^{h}_{r,t}  \leq T^{h, \text{set}}_{r,t} + 1 \qquad \qquad \ \forall \ h, \ t \\
\end{split}
\label{eq:optimization_problem}
\end{equation}

\section{Methodology}
\label{sec:methodology}

This section describes the proposed RL-based coordination strategy to solve the problem described in~\cref{sec:problem_statement}. The demand aggregator follows a business-as-usual strategy (described by~\cref{eq:bau_policy}) during normal operation. In case of a DR event, as discussed in~\cref{sec:problem_statement}, the aggregator needs to identify suitable households that can deviate from their BAU policy and dispatch them appropriately so that the aggregate power consumption follows the requested tracking signal. Closely following the work presented in~\cite{bert}, we model the aggregator’s control strategy using a hierarchical control architecture comprising an RL-based ranker for identifying suitable households and a PI control based real-time dispatcher for dispatching houses based on the ranker’s output. The real-time dispatcher sends out actions every minute, contrary to the 15 minute interval used during normal operation.

\subsection{RL-based Ranking}
As discussed in~\cref{sec:problem_statement}, the aggregator must choose households such that the energy deviation of each household due to the DR event is minimized. For this, the RL-based ranker uses the energy consumption of each household to rank different households and identify suitable houses during a DR-event. With each house modelled as an (episodic) MDP, the cumulative energy consumption over a day for a household following control policy $\pi^{h}_{b}$ can be defined using the $Q$-function (or state-action value function) as:
\begin{equation}
\begin{split}
    Q^{h}\left(\textbf{x}^{h}_{t}, u^{h}_{t}\right) &= G^{h}_t\left(\textbf{x}^{h}_{t}, \> u^{h}_t\right) + \sum_{k=1}^{k=T}G^{h}_t\left(\textbf{x}^{h}_{t+k}, \> \pi^{h}_{b}(\textbf{x}^{h}_{t+k})\right) \\
    Q^{h}\left(\textbf{x}^{h}_{t}, u^{h}_{t}\right) &= G^{h}_t\left(\textbf{x}^{h}_{t}, \> u^{h}_t\right) + Q^{h} \left(\textbf{x}^{h}_{t+1}, \>  \pi^{h}_{b}(\textbf{x}^{h}_{t+1})\right)
\end{split}
\label{eq:q_function}
\end{equation}

This $Q$-function is then used to obtain an advantage function~($A^{h}$) defined as:
\begin{equation}
    A^{h}(\textbf{x}^{h}_{t}, u^{h}_{t}) = Q^{h}(\textbf{x}^{h}_{t}, u^{h}_{t}) - V^{h} (\textbf{x}^{h}_{t})
\label{eq:adv_function}
\end{equation}
where $V^{h}$ denotes the value function and quantifies the value of being in state~$\textbf{x}^{h}_{t}$.
As discussed in~\cite{bert}, this advantage function quantifies the effect of deviating from the BAU policy for the given state $\textbf{x}^{h}_{t}$. Combining this with~\cref{eq:q_function}, we observe that, for our case, the advantage function approximates the deviation in cumulative energy consumption if an action $u^{h}_{t}$ is chosen instead of the business-as-usual action~($\pi^{h}_{b}(\textbf{x}^{h}_{t})$). Thus, for each household, the advantage value~($A^{h}(\textbf{x}^{h}_{t}, u^{h}_{t})$) provides an estimate of the ``cost'' of deviating from the BAU policy and can be used as a method for ranking different households.

To obtain these advantage functions and hence the ranks, we first need to approximate the $Q$-function for each household. This represents a standard RL problem of policy evaluation~\cite{sutton-barto} and is resolved by training $Q$-networks for each household following the fitted $Q$-iteration algorithm~\cite{fqi}. The $Q$-networks for each household are trained on a batch of previous 30 days of data ($\sim$2.8k samples) and are retrained at the beginning of each day to include the latest data.

\subsection{Real-time Dispatcher}
During a DR event, the real-time dispatcher is responsible for following the system operator tracking signal as closely as possible. For this, the dispatcher carries out three basic steps:
\begin{enumerate*}[(1)]
\item Collect advantage functions for the current states of each household;
\item Create a ranking hash table using these advantage values; and
\item Dispatch control actions for each household using this ranking table and a PI controller.
\end{enumerate*}
To ensure good signal tracking, the dispatcher sends out actions every minute, dispatching houses with the lowest rank first and gradually increasing the number of dispatched houses based on the PI controller. Additionally, to ensure that thermal comfort of users is always maintained, the dispatcher actively filters control actions, overriding actions that can lead to comfort violations~(\ie room temperature exceeding the $\pm$ 1{\centigrades} comfort band) for individual households. 

\section{Results and Discussion}
\label{sec:results}
In this section, we describe the pilot study implemented and present the results obtained. The pilot study was conducted in collaboration with anonymous company. 
This proof-of-concept study comprised 8 residential apartments that were controlled over a period of 4 weeks~(20 weekdays) during February this year~(2023). The software for the control framework described in \cref{sec:methodology} was implemented in Python and hosted on an in-premise Kubernetes cluster, interacting with the company API for sending control actions and receiving households’ states. 

The main objective of this pilot study was to demonstrate a scalable, data-driven RL-based coordination mechanism for residential households without using external simulators or building models. This includes answering two main questions:
\begin{enumerate*}[(Q1)]
\item How closely can the aggregate energy consumption of the cluster follow the tracking signal?
\item Does the RL-based ranking system learn any useful patterns to select households as compared to random selection?
\end{enumerate*}
The following results aim to address these two questions.

\subsection{Overall Response}
To address Q1 regarding the signal tracking capacity of our control framework, we set a square wave DR signal of 40~minutes duration~(twice per day) and observed the behavior of the system over 10~days. 

\begin{figure}
    \centering
    \includegraphics[width=.85\columnwidth]{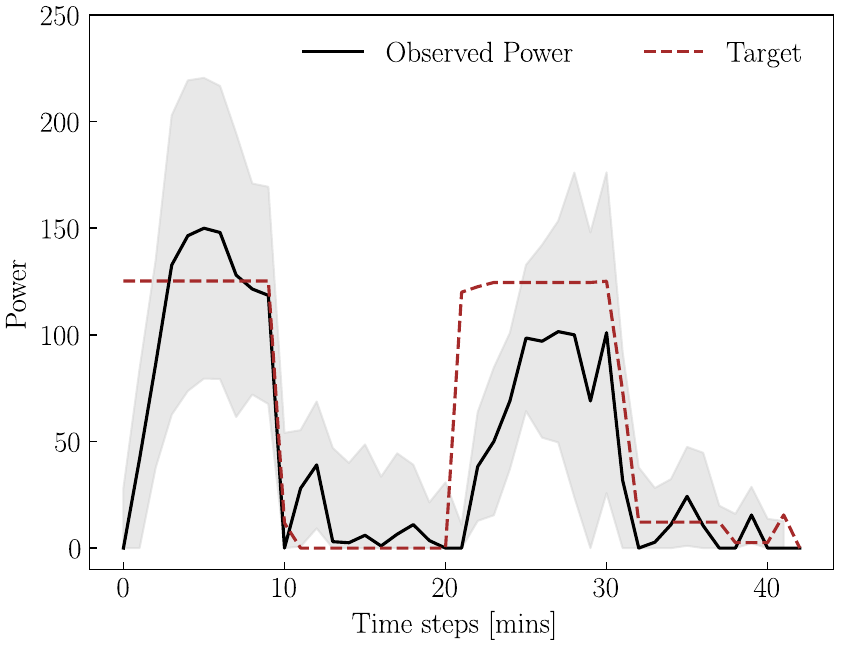}
    \caption{Consolidated performance of DR events. The solid lines represent the median values for observed aggregated power and the shaded area represents the standard deviation across all DR events performed. The target values represent the DR signals provided to the aggregator for tracking.}
    \label{fig:overall_response}
\end{figure}

\Cref{fig:overall_response} presents consolidated results over the 10 days.
It can be observed that the median value of aggregate power of the entire cluster is able to follow the trend of the square wave tracking signal. However, it falls short of accurately following the signal, especially during the second peak. This is due to the lack of flexibility available during the DR event and can be largely contributed to the small number of households~(8) available in this case study. However, while the small number of households limit the accuracy, the overall trend demonstrates that our control framework can steer the aggregate energy consumption of the cluster to satisfactorily follow an external tracking signal.

\subsection{Quality of RL-based Ranking}
\begin{figure}
    \centering
    \includegraphics[width=0.85\columnwidth]{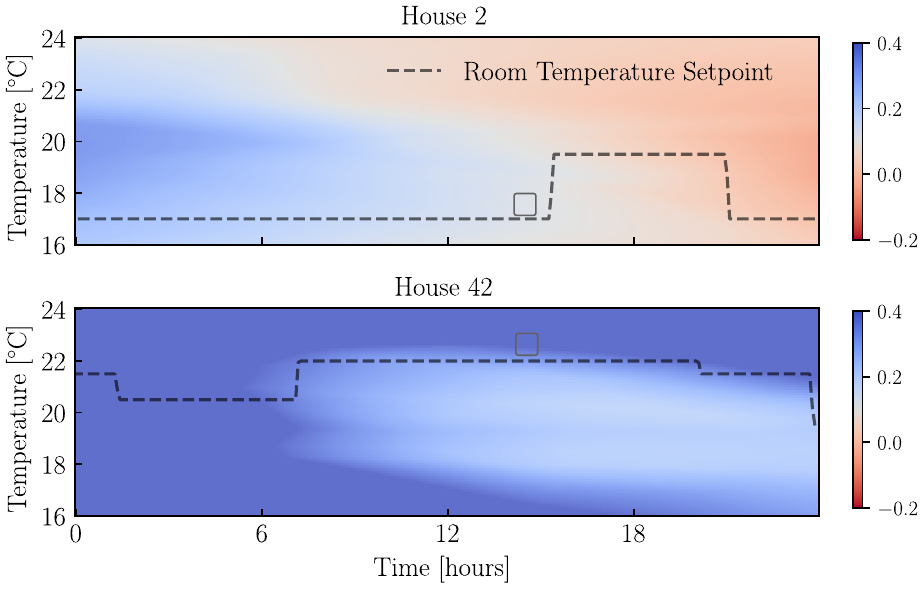}
\caption{Heatmaps depicting the `dis'-advantage of switching ON over switching OFF action~(\ie $A^{h}(\textbf{x}^{h}_{t}, 1) - A^{h}(\textbf{x}^{h}_{t},0)$). Blue regions indicate that the $Q$-network believes it is better to remain switched off or switching ON is expected to be more costly (in terms of energy consumption). Conversely, red regions imply a big advantage to switching ON. The dashed line shows the user-defined temperature setpoint for the given day and the grey box denotes the state when the DR event corresponding to~\cref{tab:ranks} occurred.}
\label{fig:q-maps}
\end{figure}

To analyze what patterns the RL-based ranking system has learnt~(\ie Q2), we conducted a smaller experiment using only 2 households.
The idea was to observe the relevance of ranks obtained in this reduced size test and provide an intuition about the RL-based based ranks.
The underlying advantage functions for both households are visualized as a heatmap in \cref{fig:q-maps}.
The heatmap presents the difference between the advantage values for switching ON~($u_t=1$) and switching OFF~($u_t=0$) for a given state~($\textbf{x}_{t}$).
The time and room temperature for state~($\textbf{x}_t$) are depicted by the grey box, which is the exact time when a DR event was activated.
Through this visualization, we observe that the heatmap for house 2 shows a clear pattern that is strongly correlated to the user-defined room temperature setpoint.
Specifically, the $Q$-network estimates that `switching ON' at any time before 15h is not advantageous and is depicted by darker shades of blue. Similarly, around 17h, the $Q$-network for house 2 expects a change in user-defined temperature setpoint and hence values switching ON almost equal to or sometimes lower than switching OFF, indicated by the warmer shades in the heatmap. This trend corresponds to the regular pattern of room temperature setpoints observed for this household. Contrary to this, for house 42, we observe a significantly different trend, where the $Q$-network learns an almost flat trend of always estimating a higher cost for switching ON (compared to switching OFF) unless the room temperature falls below a certain value (20{\centigrades}) corresponding to the mean value of past room temperature setpoint. This trend also correlates to a relatively flat room temperature setpoints set by users of this house. Note that the room temperature setpoint is a user-defined quantity which is not used as input for training the $Q$-network. The trends in~\cref{fig:q-maps} indicate that the $Q$-networks are able to learn these user-specific patterns and assign ranks based on such patterns. 

\begin{table}[t]
    \caption{Ranks assigned during a two-household upwards DR event. The action values represent non-BAU control actions.}
    \centering    
    \begin{tabular}{cccc}
    \toprule
         & Action & Advantage & Rank \\
     \midrule
     House 2    & ON     & 0.0963    & 1    \\
     House 42   & ON     & 0.3460    & 2    \\    
    \bottomrule
    \end{tabular}
    \label{tab:ranks}
\end{table}
Further, \cref{tab:ranks} shows the ranks for both these houses for an upward DR event, requiring an increase in power consumption, occurring at around 15h.
Using \cref{fig:q-maps}, these ranks can be justified since house~2 is expected to switch ON heating in any case and hence it is ranked lower than the household that is expected to keep their heating OFF for the foreseeable future. Based on these ranks, the real-time dispatcher will first switch ON heating for House 2, and if insufficient, it will move forward to house 42.  
These results indicate that the RL-based ranking system learns useful patterns related to the user-behavior and assigns ranks that are intuitive and better than a random ranking strategy. 

\section{Conclusion}
\label{sec:conclusions}

This work presented a real-world, proof-of-concept demonstration of a scalable, data-driven RL-based control strategy for coordinating the power consumption of a group of residential buildings to provide demand response services.
As presented in~\cref{sec:results}, our proposed coordination strategy shows good power tracking capacity with the RL-based ranker producing intuitive and reasonable ranks by solely relying on observed data from different households.
The aim of this work was to deliver a proof-of-concept feasibility study and hence we focused on a relatively small set of households.
While this small size had practical benefits including ease of deployment, it has led to a few limitations that we aim to address in future work.

\subsection{Limitations and Future Work}
As discussed earlier, our current work presented a pilot study of a relatively small size (in terms of both the number of households as well as the deployment period). 
While this had practical benefits, future work would involve expanding the size of this study by adding more households and testing the control strategy over an entire winter season. This would allow testing the proposed coordination strategy for tracking arbitrary signals and analyzing the commercial potential of this technique.
Besides the size of the study, we have identified two key areas for improvement. The first relates to the need for a validation and backtesting framework for the trained $Q$-functions and advantage functions. Future work will research policy evaluation methods~\cite{offpolicy_eval} and data-driven, physics informed simulators~\cite{physnet} for data-driven backtesting strategies.
The second improvement area involves using transfer learning-based methods to leverage data from different households to pre-train base models that can be finetuned in a data-efficient way to obtain the necessary $Q$-functions.
This will ensure robust training performance and allow us to use advanced deep learning architectures such as transformers that can improve the overall accuracy and scalability of the learnt $Q$-functions.

\section{Acknowledgements}
This pilot study was performed in collaboration with domX --- a Greek IoT-based energy management company --- and received funding through the European Union's Horizon 2020 Projects BiGG~(grant no.\ 957047) and 
BRIGHT~(grant no.\ 957816).
We thank Polychronis Symeonidis, Stratos Kerandis and the entire domX team for their support throughout this study. We also thank Matthias Strobbe for his support during this study. 


\bibliographystyle{unsrtnat}
\bibliography{domx_pilot_refs}

\begin{thebibliography}{26}
\providecommand{\natexlab}[1]{#1}
\providecommand{\url}[1]{\texttt{#1}}
\expandafter\ifx\csname urlstyle\endcsname\relax
  \providecommand{\doi}[1]{doi: #1}\else
  \providecommand{\doi}{doi: \begingroup \urlstyle{rm}\Url}\fi

\bibitem[Doe(2015)]{built_env_con}
U~Doe.
\newblock Chapter 5: Increasing efficiency of building systems and technologies.
\newblock \emph{Quadrennial Technology Review: An Assessment of Energy Technologies and Research Opportunities}, pages 143--181, 2015.

\bibitem[Zhang et~al.(2022)Zhang, Seal, Wu, Bouffard, and Boulet]{mpc_rl_review}
Huiliang Zhang, Sayani Seal, Di~Wu, François Bouffard, and Benoit Boulet.
\newblock Building {Energy} {Management} {With} {Reinforcement} {Learning} and {Model} {Predictive} {Control}: {A} {Survey}.
\newblock \emph{IEEE Access}, 10:\penalty0 27853--27862, 2022.
\newblock ISSN 2169-3536.
\newblock \doi{10.1109/ACCESS.2022.3156581}.

\bibitem[Li et~al.(2022)Li, Satchwell, Finn, Christensen, Kummert, Le~Dréau, Lopes, Madsen, Salom, Henze, and Wittchen]{energy_flex}
Rongling Li, Andrew~J. Satchwell, Donal Finn, Toke~Haunstrup Christensen, Michaël Kummert, Jérôme Le~Dréau, Rui~Amaral Lopes, Henrik Madsen, Jaume Salom, Gregor Henze, and Kim Wittchen.
\newblock Ten questions concerning energy flexibility in buildings.
\newblock \emph{Building and Environment}, 223:\penalty0 109461, September 2022.
\newblock ISSN 0360-1323.
\newblock \doi{10.1016/j.buildenv.2022.109461}.
\newblock URL \url{https://www.sciencedirect.com/science/article/pii/S0360132322006928}.

\bibitem[Yao and Shekhar(2021)]{mpc_review_2023}
Ye~Yao and Divyanshu~Kumar Shekhar.
\newblock State of the art review on model predictive control ({MPC}) in {Heating} {Ventilation} and {Air}-conditioning ({HVAC}) field.
\newblock \emph{Building and Environment}, 200:\penalty0 107952, August 2021.
\newblock ISSN 03601323.
\newblock \doi{10.1016/j.buildenv.2021.107952}.
\newblock URL \url{https://linkinghub.elsevier.com/retrieve/pii/S0360132321003565}.

\bibitem[Fu et~al.(2022)Fu, Han, Chen, Lu, Wu, and Wang]{rl_review}
Qiming Fu, Zhicong Han, Jianping Chen, You Lu, Hongjie Wu, and Yunzhe Wang.
\newblock Applications of reinforcement learning for building energy efficiency control: {A} review.
\newblock \emph{Journal of Building Engineering}, 50:\penalty0 104165, June 2022.
\newblock ISSN 2352-7102.
\newblock \doi{10.1016/j.jobe.2022.104165}.
\newblock URL \url{https://www.sciencedirect.com/science/article/pii/S2352710222001784}.

\bibitem[Drgoňa et~al.(2020)Drgoňa, Arroyo, Cupeiro~Figueroa, Blum, Arendt, Kim, Ollé, Oravec, Wetter, Vrabie, and Helsen]{mpc-basics}
Ján Drgoňa, Javier Arroyo, Iago Cupeiro~Figueroa, David Blum, Krzysztof Arendt, Donghun Kim, Enric~Perarnau Ollé, Juraj Oravec, Michael Wetter, Draguna~L. Vrabie, and Lieve Helsen.
\newblock All you need to know about model predictive control for buildings.
\newblock \emph{Annual Reviews in Control}, 50:\penalty0 190--232, 2020.
\newblock ISSN 13675788.
\newblock \doi{10.1016/j.arcontrol.2020.09.001}.
\newblock URL \url{https://linkinghub.elsevier.com/retrieve/pii/S1367578820300584}.

\bibitem[Gorecki et~al.(2017)Gorecki, Fabietti, Qureshi, and Jones]{swiss_mpc_frequency}
Tomasz~T. Gorecki, Luca Fabietti, Faran~A. Qureshi, and Colin~N. Jones.
\newblock Experimental demonstration of buildings providing frequency regulation services in the {Swiss} market.
\newblock \emph{Energy and Buildings}, 144:\penalty0 229--240, June 2017.
\newblock ISSN 03787788.
\newblock \doi{10.1016/j.enbuild.2017.02.050}.
\newblock URL \url{https://linkinghub.elsevier.com/retrieve/pii/S0378778816311616}.

\bibitem[Wang et~al.(2023)Wang, Chen, Wang, Gao, and Wang]{wang_mpc_field_2023}
Dan Wang, Yangzhe Chen, Wei Wang, Cheng Gao, and Zhe Wang.
\newblock Field test of {Model} {Predictive} {Control} in residential buildings for utility cost savings.
\newblock \emph{Energy and Buildings}, 288:\penalty0 113026, June 2023.
\newblock ISSN 0378-7788.
\newblock \doi{10.1016/j.enbuild.2023.113026}.
\newblock URL \url{https://www.sciencedirect.com/science/article/pii/S0378778823002566}.

\bibitem[De~Coninck and Helsen(2016)]{mpc_brussels}
Roel De~Coninck and Lieve Helsen.
\newblock Practical implementation and evaluation of model predictive control for an office building in {Brussels}.
\newblock \emph{Energy and Buildings}, 111:\penalty0 290--298, January 2016.
\newblock ISSN 0378-7788.
\newblock \doi{10.1016/j.enbuild.2015.11.014}.
\newblock URL \url{https://www.sciencedirect.com/science/article/pii/S0378778815303790}.

\bibitem[Azuatalam et~al.(2020)Azuatalam, Lee, de~Nijs, and Liebman]{rl_hvac_2020}
Donald Azuatalam, Wee-Lih Lee, Frits de~Nijs, and Ariel Liebman.
\newblock Reinforcement learning for whole-building {HVAC} control and demand response.
\newblock \emph{Energy and AI}, 2:\penalty0 100020, November 2020.
\newblock ISSN 2666-5468.
\newblock \doi{10.1016/j.egyai.2020.100020}.
\newblock URL \url{https://www.sciencedirect.com/science/article/pii/S2666546820300203}.

\bibitem[Zhang et~al.(2019)Zhang, Chong, Pan, Zhang, and Lam]{rl_real_1}
Zhiang Zhang, Adrian Chong, Yuqi Pan, Chenlu Zhang, and Khee~Poh Lam.
\newblock Whole building energy model for {HVAC} optimal control: {A} practical framework based on deep reinforcement learning.
\newblock \emph{Energy and Buildings}, 199:\penalty0 472--490, September 2019.
\newblock ISSN 0378-7788.
\newblock \doi{10.1016/j.enbuild.2019.07.029}.
\newblock URL \url{https://www.sciencedirect.com/science/article/pii/S0378778818330858}.

\bibitem[Coraci et~al.(2023)Coraci, Brandi, Hong, and Capozzoli]{tl_rl_buildings}
Davide Coraci, Silvio Brandi, Tianzhen Hong, and Alfonso Capozzoli.
\newblock Online transfer learning strategy for enhancing the scalability and deployment of deep reinforcement learning control in smart buildings.
\newblock \emph{Applied Energy}, 333:\penalty0 120598, March 2023.
\newblock ISSN 0306-2619.
\newblock \doi{10.1016/j.apenergy.2022.120598}.
\newblock URL \url{https://www.sciencedirect.com/science/article/pii/S0306261922018554}.

\bibitem[Nweye et~al.(2023)Nweye, Sankaranarayanan, and Nagy]{merlin}
Kingsley Nweye, Siva Sankaranarayanan, and Zoltan Nagy.
\newblock Merlin: Multi-agent offline and transfer learning for occupant-centric operation of grid-interactive communities.
\newblock \emph{Applied Energy}, 346:\penalty0 121323, 2023.
\newblock \doi{https://doi.org/10.1016/j.apenergy.2023.121323}.

\bibitem[Pinto et~al.(2021)Pinto, Piscitelli, V{\'a}zquez-Canteli, Nagy, and Capozzoli]{coordinated_DRL}
Giuseppe Pinto, Marco~Savino Piscitelli, Jos{\'e}~Ram{\'o}n V{\'a}zquez-Canteli, Zolt{\'a}n Nagy, and Alfonso Capozzoli.
\newblock Coordinated energy management for a cluster of buildings through deep reinforcement learning.
\newblock \emph{Energy}, 229:\penalty0 120725, 2021.
\newblock \doi{https://doi.org/10.1016/j.energy.2021.120725}.

\bibitem[Xie et~al.(2023)Xie, Ajagekar, and You]{multi-agent_2023}
Jiahan Xie, Akshay Ajagekar, and Fengqi You.
\newblock Multi-{Agent} attention-based deep reinforcement learning for demand response in grid-responsive buildings.
\newblock \emph{Applied Energy}, 342:\penalty0 121162, July 2023.
\newblock ISSN 0306-2619.
\newblock \doi{10.1016/j.apenergy.2023.121162}.
\newblock URL \url{https://www.sciencedirect.com/science/article/pii/S0306261923005263}.

\bibitem[Vazquez-Canteli et~al.(2020)Vazquez-Canteli, Henze, and Nagy]{marlisa}
Jose~R Vazquez-Canteli, Gregor Henze, and Zoltan Nagy.
\newblock Marlisa: Multi-agent reinforcement learning with iterative sequential action selection for load shaping of grid-interactive connected buildings.
\newblock In \emph{Proceedings of the 7th ACM international conference on systems for energy-efficient buildings, cities, and transportation}, pages 170--179, 2020.
\newblock \doi{https://doi.org/10.1145/3408308.3427604}.

\bibitem[Blum et~al.(2021)Blum, Arroyo, Huang, Drgo{\v{n}}a, Jorissen, Walnum, Chen, Benne, Vrabie, Wetter, et~al.]{boptest}
David Blum, Javier Arroyo, Sen Huang, J{\'a}n Drgo{\v{n}}a, Filip Jorissen, Harald~Taxt Walnum, Yan Chen, Kyle Benne, Draguna Vrabie, Michael Wetter, et~al.
\newblock Building optimization testing framework (boptest) for simulation-based benchmarking of control strategies in buildings.
\newblock \emph{Journal of Building Performance Simulation}, 14\penalty0 (5):\penalty0 586--610, 2021.
\newblock \doi{https://doi.org/10.1080/19401493.2021.1986574}.

\bibitem[V\'{a}zquez-Canteli et~al.(2019)V\'{a}zquez-Canteli, K\"{a}mpf, Henze, and Nagy]{citylearn}
Jos\'{e}~R. V\'{a}zquez-Canteli, J\'{e}r\^{o}me K\"{a}mpf, Gregor Henze, and Zoltan Nagy.
\newblock Citylearn v1.0: An openai gym environment for demand response with deep reinforcement learning.
\newblock In \emph{Proceedings of the 6th ACM International Conference on Systems for Energy-Efficient Buildings, Cities, and Transportation}, BuildSys '19, page 356–357, New York, NY, USA, 2019. Association for Computing Machinery.
\newblock ISBN 9781450370059.
\newblock \doi{10.1145/3360322.3360998}.
\newblock URL \url{https://www.citylearn.net/index.html}.

\bibitem[Crawley et~al.(2001)Crawley, Lawrie, Winkelmann, Buhl, Huang, Pedersen, Strand, Liesen, Fisher, Witte, et~al.]{energyplus}
Drury~B Crawley, Linda~K Lawrie, Frederick~C Winkelmann, Walter~F Buhl, Y~Joe Huang, Curtis~O Pedersen, Richard~K Strand, Richard~J Liesen, Daniel~E Fisher, Michael~J Witte, et~al.
\newblock Energyplus: creating a new-generation building energy simulation program.
\newblock \emph{Energy and buildings}, 33\penalty0 (4):\penalty0 319--331, 2001.
\newblock \doi{https://doi.org/10.1016/S0378-7788(00)00114-6}.

\bibitem[Wetter et~al.(2014)Wetter, Zuo, Nouidui, and Pang]{modelica}
Michael Wetter, Wangda Zuo, Thierry~S Nouidui, and Xiufeng Pang.
\newblock Modelica buildings library.
\newblock \emph{Journal of Building Performance Simulation}, 7\penalty0 (4):\penalty0 253--270, 2014.
\newblock \doi{https://doi.org/10.1080/19401493.2013.765506}.

\bibitem[Reynders et~al.(2018)Reynders, Lopes, Marszal-Pomianowska, Aelenei, Martins, and Saelens]{thermal_mass}
Glenn Reynders, Rui~Amaral Lopes, Anna Marszal-Pomianowska, Daniel Aelenei, Jo{\~a}o Martins, and Dirk Saelens.
\newblock Energy flexible buildings: An evaluation of definitions and quantification methodologies applied to thermal storage.
\newblock \emph{Energy and Buildings}, 166:\penalty0 372--390, 2018.
\newblock \doi{https://doi.org/10.1016/j.enbuild.2018.02.040}.

\bibitem[Sutton and Barto(2018)]{sutton-barto}
Richard~S Sutton and Andrew~G Barto.
\newblock \emph{Reinforcement learning: An introduction}.
\newblock MIT press, 2018.

\bibitem[Liu et~al.(2019)Liu, Peeters, Callaway, and Claessens]{bert}
Mingxi Liu, Stef Peeters, Duncan~S. Callaway, and Bert~J. Claessens.
\newblock Trajectory {Tracking} {With} an {Aggregation} of {Domestic} {Hot} {Water} {Heaters}: {Combining} {Model}-{Based} and {Model}-{Free} {Control} in a {Commercial} {Deployment}.
\newblock \emph{IEEE Transactions on Smart Grid}, 10\penalty0 (5):\penalty0 5686--5695, September 2019.
\newblock ISSN 1949-3061.
\newblock \doi{10.1109/TSG.2018.2890275}.

\bibitem[Ernst et~al.(2005)Ernst, Geurts, and Wehenkel]{fqi}
Damien Ernst, Pierre Geurts, and Louis Wehenkel.
\newblock Tree-based batch mode reinforcement learning.
\newblock \emph{Journal of Machine Learning Research}, 6, 2005.

\bibitem[Chen et~al.(2020)Chen, Jin, Wang, Hong, and Berg{\'e}s]{offpolicy_eval}
Bingqing Chen, Ming Jin, Zhe Wang, Tianzhen Hong, and Mario Berg{\'e}s.
\newblock Towards off-policy evaluation as a prerequisite for real-world reinforcement learning in building control.
\newblock In \emph{Proceedings of the 1st International Workshop on Reinforcement Learning for Energy Management in Buildings \& Cities}, pages 52--56, 2020.

\bibitem[Gokhale et~al.(2022)Gokhale, Claessens, and Develder]{physnet}
Gargya Gokhale, Bert Claessens, and Chris Develder.
\newblock Physics informed neural networks for control oriented thermal modeling of buildings.
\newblock \emph{Applied Energy}, 314:\penalty0 118852, 2022.
\newblock \doi{https://doi.org/10.1016/j.apenergy.2022.118852}.

\end{thebibliography}

\end{document}